\documentclass[prb,10pt,notitlepage,twocolumn,superscriptaddress]{revtex4}

\usepackage{graphicx}
\usepackage{amsmath}
\usepackage{amssymb}
\usepackage{amsfonts}
\usepackage{color}
\usepackage{xspace}
\usepackage{subfigure}
\usepackage{bm}
\usepackage{bbold}

\newcommand{\moly}{Mo$_3$S$_7$(dmit)$_3$\xspace}
\newcommand{\bra}[1]{\langle #1 |}
\newcommand{\ket}[1]{| #1 \rangle}

\renewcommand{\cite}[1]{[\onlinecite{#1}]}

\begin{document}

\title{Spin-orbit coupling in {\moly}}
\author{A. C. Jacko }\affiliation{School of Mathematics and Physics, The University of Queensland, Brisbane, Queensland, 4072, Australia}
\author{A. L. Khosla}\affiliation{School of Mathematics and Physics, The University of Queensland, Brisbane, Queensland, 4072, Australia}
\author{J. Merino}\affiliation{Departmento de F\'{i}sica Te\'{o}rica de la Materia Condensada, Condensed Matter Physics Centre (IFIMAC) and Instituto Nicol\'{a}s Cabrera, Universidad Aut\'{o}noma de Madrid, Madrid 28049, Spain}
\author{B. J. Powell}\affiliation{School of Mathematics and Physics, The University of Queensland, Brisbane, Queensland, 4072, Australia}

\begin{abstract}
Spin-orbit coupling in crystals is known to lead to unusual direction dependent exchange interactions, however understanding of the consequeces of such effects in molecular crystals is incomplete.
Here we perform four component relativistic density functional theory computations on the multi-nuclear molecular crystal {\moly} and show that both intra- and inter-molecular spin-orbit coupling are significant. 
We determine a long-range relativistic single electron Hamiltonian from first principles by constructing Wannier spin-orbitals. We analyse the various contributions through the lens of group theory. 
Intermolecular spin-orbit couplings like those found here are known to lead to quantum spin-Hall and topological insulator phases on the 2D lattice formed by the tight-binding model predicted for a single layer of {\moly}.
\end{abstract}

\maketitle
\section{Introduction}

The interplay of spin-orbit coupling (SOC) and strong electronic correlations is an important theme in condensed matter physics. 
The competition between these mechanisms leads to many novel phases of matter with exciting technological properties, including spin liquid phases \cite{jackeli09,galitski10,pesin10,balents10,brink15,kee16,balents14}.
To date, the primary focus of this research has been on transition metal oxides \cite{kee16,balents14,jackeli09}. 
However, SOC is known to be significant in organic and organometallic materials, where many parameters are tunable by chemical substitutions \cite{balents10,winter15,smith04,powell2015}. 
In such organometallic molecules, SOC is typically treated as a property of single atoms.

Only a handful of materials have been identified as candidate spin liquids, \cite{lee08,balents10,powell11,isono14} and there is some suggestion that {\moly} could be among them \cite{janani14a,noursehaldane}.
{\moly} is a single component organometallic molecular insulator with localized magnetic moments, but no long-range magnetic order \cite{llusar04}. It has been seen to have a layered tight-binding lattice \cite{jacko15a}.

\begin{figure}
\begin{center}
\includegraphics[width=0.6\columnwidth]{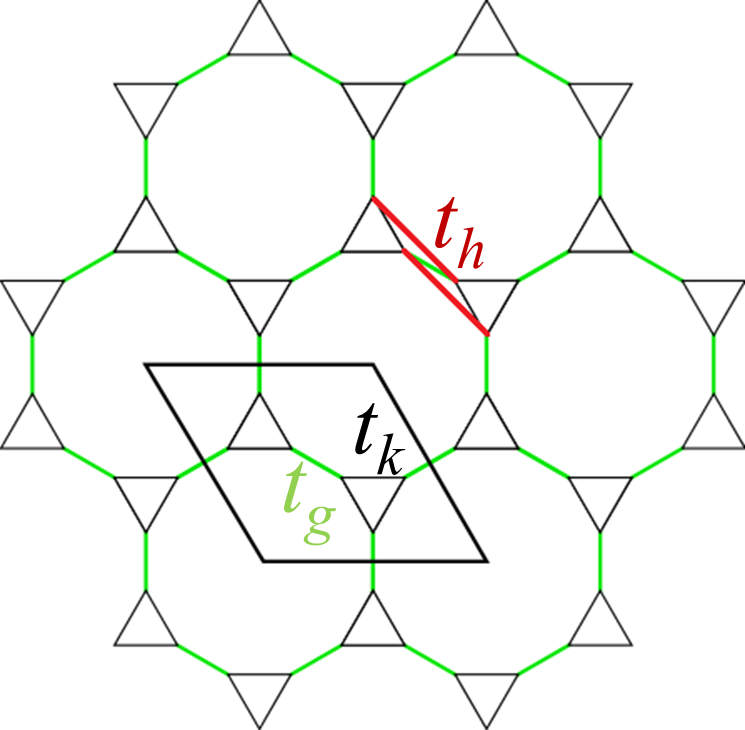} 
\caption{The two-dimensional lattice of {\moly}, known as the decorated honeycomb lattice, the star lattice, and the kagomene lattice; the latter since it interpolates between the kagom\'e lattice and the honeycomb lattice. The kagom\'e-like hopping (black) is labeled $t_k$, while the graphene-like hopping (green) is labeled $t_g$. The full 3D lattice stacks layers of the kagomene lattice directly on top of one another in the $z$ direction. 
An example of the in-plane next-nearest neighbour hopping $t_h$ (red) is also indicated; this hopping is chiral (it preserves inversion symmetry between the pair of molecules while breaking the reflection symmetry, as there is no reflection plane in the crystal).}\label{fig:kagomene}
\end{center}
\end{figure}

The layered lattice of {\moly} (Fig. \ref{fig:kagomene}) leads to Dirac points, quasi-1D bands, and topological states \cite{jacko15a}.
This lattice (known variously as the decorated honeycomb lattice, the star lattice, and the kagomene lattice) supports topological insulating phases and a quantum spin-Hall effect when spin-orbit coupling is included \cite{ruegg10,wen10}. 
We recently showed that $\mathcal{C}_N$ molecules have a spin-molecular orbital coupling (SMOC) due to orbital currents around the molecules, which can lead to anisotropic and direction dependent exchange interactions \cite{khosla16Cn}.  
These anisotropic exchange interactions can lead to the physical realisation of compass models (the most studied of which is the spin-$\frac{1}{2}$ Kitaev model) \cite{kitaev06,jackeli09,nussinov15,merino16moly}.
Here we show that inter-molecular SOC is significant in {\moly} and should not be neglected. 
Multi-nuclear organometallic complexes thus have all of the required features to realise compass models. The chemical modifications possible in these materials provide an avenue for tuning the parameters of such models to enhance these effects.

Here we report a powerful demonstration of the Wannier orbital construction technique - the determination of first principles intra- and inter-molecular spin-orbit coupling parameters for {\moly} from a four component relativistic calculation. These parameters come naturally from the computation of a first principles Hamiltonian in the Wannier basis.
The Wannier orbital (WO) overlaps in this relativistic calculation include both regular hopping terms and spin-orbit coupling terms. 
The largest effects of relativity are captured by a simple model of molecular angular momentum states (analogous to the usual treatment of atomic angular momentum states). Both intra- and inter-molecular spin-orbit coupling overlaps are present, and may play an important role in determining the ground state properties of {\moly}.
By applying this first principles relativistic parameterization proceedure we can better understand the path to designing compass models in molecular crystals.

\subsection{SMOC in $\mathcal{C}_3$ complexes}

\begin{figure}
\begin{center}
\includegraphics[width=0.6\columnwidth]{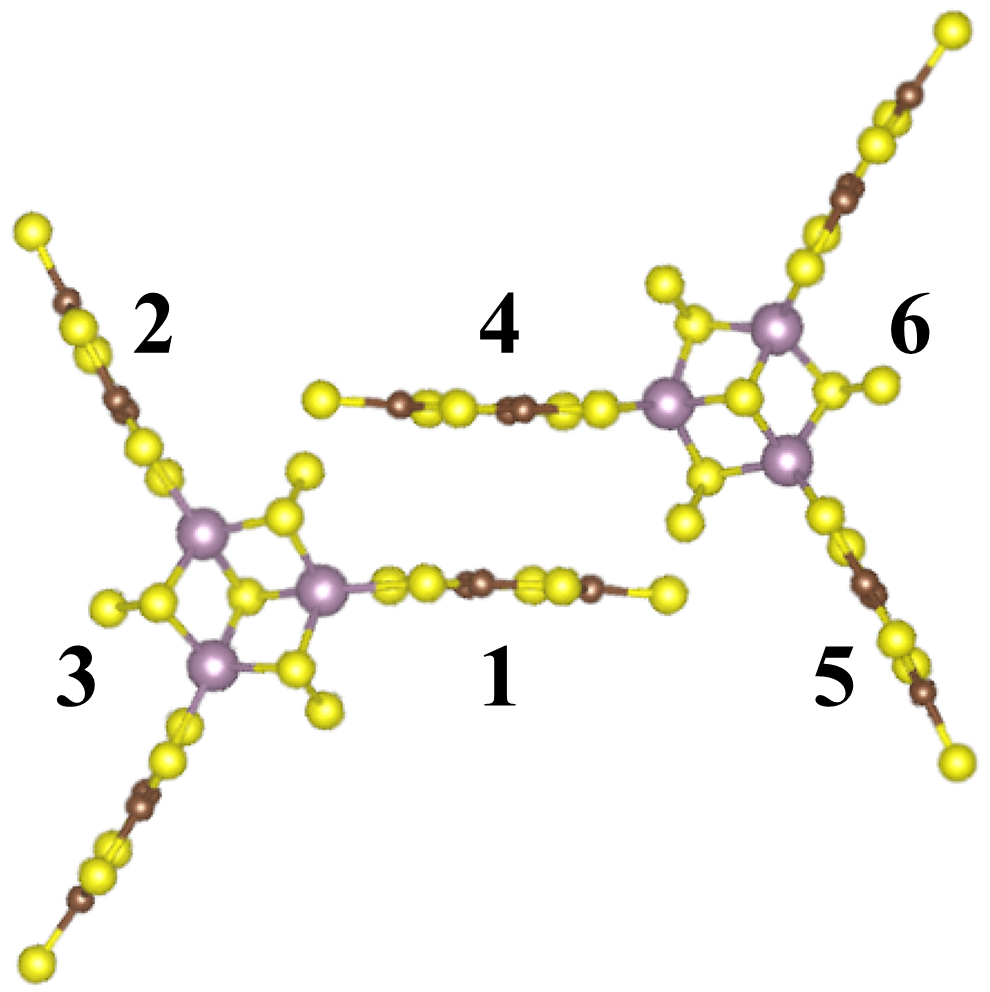} 
\caption{A pair {\moly} molecules, with molybdenum atoms in purple, carbon in brown, and sulphur in yellow. The canting of the dmit ligands and the vertical asymmetry of the Mo$_3$S$_7$ core means that this molecule has $\mathcal{C}_3$ symmetry; it is symmetric only under rotations by $2\pi / 3$. The pair of molecules are related by an inversion centre, mapping site $i$ to $i+3$.}\label{fig:molyview}
\end{center}
\end{figure}

In the absence of SOC, the first principles Hamiltonian of {\moly} (Fig. \ref{fig:molyview}) is a layered `kagomene' lattice; each molecule is a triangular ring of sites, connected to each other on a stacked chiral honeycomb lattice \cite{jacko15a}. It is thus an example of a $\mathcal{C}_3$ complex, for which the form of the SMOC Hamiltonian is known \cite{khosla16Cn}. We breifly review that result.
The tight-binding Hamiltonian of the (ring-like) coordination complex is
\[
\hat{H}^{c}_0 = -t_k \sum_{j,\sigma} c^\dagger_{j \sigma} c_{j+1 \sigma} + h.c.,
\]
where $j$ is an integer labeling the position of each site around the ring-like complex, $c^\dagger_j$ creates an electron on site $j$. In the case of {\moly}, each `site' is a local Wannier orbital with weight on one of the dmit ligands plus the molybdenum-sulfur core.
This Hamiltonian can be diagonalised by transforming to a Bloch (plane wave)-like basis on the ring, $c^\dagger_{q\sigma} = i^{|q|}\sum_{j=1}^{3} e^{i \phi q (j-1)} c^\dagger_{j\sigma} / \sqrt{3}$,
with eigenvalues $E_q = -2 t_k \cos (\phi q)$, and $\phi = 2 \pi / 3$. The prefactor $i^{|q|}$ ensures that these states transform as angular momentum under time reversal.
Since these molecular states are on a ring, this Bloch-like momentum is equivalent to a molecular orbital angular momentum; in this case an $L_{mol}=1, L_{mol}^z= \{ -1, 0, 1\}$ set of states (much like atomic $p$-orbitals). It is worth noting that the analogy with atomic (spherically symmetric) spin-orbit coupling only holds for $\mathcal{C}_N$ molecules with odd-$N$. For even-$N$, $L_{mol}$ is non-integer, and $\ket{L_z^{max}} = \ket{L_z^{min}}$ \cite{khosla16Cn}.

The molecular orbital angular momentum leads to a spin-orbit coupling interaction analagous to that in atomic orbital angular momentum states.
For this system with $\mathcal{C}_3$ symmetry, the spin-orbit coupling operator is \cite{khosla16Cn}
\begin{equation}
\hat{H}_{SMO} = \lambda^z L^z_{mol} S^z + \frac{\lambda^{xy}}{2}\left( L^+_{mol} S^- + L^-_{mol} S^+ \right). \label{eq:Hsoc}
\end{equation}
where
\[
L^z_{mol} = \sum_{\nu, \sigma} \nu c^\dagger_{\nu \sigma} c_{\nu \sigma}
\]
with $\nu \in \{1,0,-1\}$, and
\[
L^+_{mol} = \sum_\sigma  c^\dagger_{1 \sigma} c_{0 \sigma} + c^\dagger_{0 \sigma} c_{-1 \sigma},
\]
\[
L^-_{mol} = \sum_\sigma  c^\dagger_{-1 \sigma} c_{0 \sigma} + c^\dagger_{0 \sigma} c_{1 \sigma}.
\]
If $\lambda^z = \lambda^{xy} = \lambda$ (the spherically symmetric case), then $\hat{H}_{SMO} = \lambda \bm{L}_{mol} \cdot \bm{S}$.

Thus we have a molecular Hamiltonian that includes spin orbit coupling, $\hat{H}^{c} = \hat{H}^{c}_0 + \hat{H}_{SMO}$. 
We now embed this molecular model into the full lattice structure and compare it to four component DFT computations.

\section{Four Component Relativistic Density Functional Theory}

An \textit{ab initio} Hamiltonian for {\moly} has been constructed previously by producing localized Wannier orbitals from a DFT computation without spin-orbit coupling \cite{jacko15a}. This approach is particularly well suited to organic and organo-metallic molecular crystals due to the separation of energy scales in this class of material \cite{nakamura12,marzari12,jacko13dmit,jackounp}.
Previous calculations on {\moly} included only scalar relativistic effects; here we report a more intensive computation that includes a full four-component representation of the effects of relativity \cite{eschrig04}. We performed four-component ``full-relativistic'' DFT calculations in an all-electron full-potential local orbital basis using the FPLO package \cite{koepernik99}; the density was converged on a $(8 \times 8 \times 8)$ $k$ mesh using the PBE generalized gradient approximation \cite{perdew96}. This four component calcuation includes complex Dirac spinor fields, allowing for a more complete treatment of relativity than via a scalar correction.
Since the four-component calculation includes spin-orbit coupling, we must treat each spin explicitly; we need separate Wannier orbitals for each spin.
Localized WOs were constructed from the twelve spinful bands closest to the Fermi energy, and real-space overlaps were computed to construct an \textit{ab initio} single electron Hamiltonian. 
The complex overlaps between the Wannier orbitals produced a spin-dependent model Hamiltonian that includes tight-binding and relativistic effects.

\begin{figure}
\begin{center}
\includegraphics[width=0.9\columnwidth]{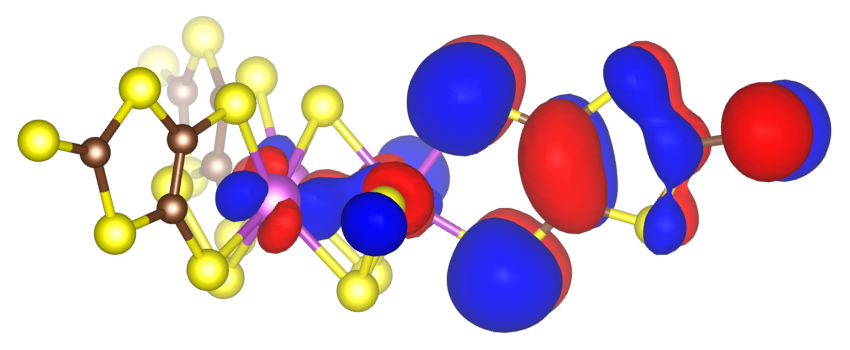} 
\caption{Twelve Wannier spin-orbitals of {\moly} determine its low energy physics. The three Kramers pairs per molecule are related to each other by the $\tilde{\cal{C}}_3$ symmetry of the molecule, and the two molecules per unit cell are related by an inversion centre between them. The real part of the spin-up component of one of the spin-up Wannier orbitals is shown (the other components are orders of magnitude smaller).}\label{fig:WF4comp}
\end{center}
\end{figure}

Four component relativistic computations mix together spin-up with spin-down, and the `large' components of the Dirac spinor with the `small'. Thus the Wannier functions are not simply complex scalar fields, they are complex four-vector fields. However, as one might hope, the `small' components are orders of magnitude smaller than the dominant `large' component, and only one of spin-up or spin-down is significant in each orbital, as illustrated in Fig. \ref{fig:WF4comp}. Thus we can label the resulting Wannier orbitals with specific spin labels. In this way we have constructed a localized basis of spin orbitals to use as a basis for constructing a first principles Hamiltonian for {\moly}.

\subsection{Complex hopping parameters from spin-dependent Wannier Orbitals}

The first principles single-electron Hamiltonian for {\moly} is
\begin{equation}
\hat{H}_{rel} = \sum_{i,j} \sum_{\alpha,\beta} c^\dagger_{i\alpha} \Big(-t_{ij}  \delta_{\alpha \beta} + i \bm{\lambda}_{ij}\cdot\bm{\sigma}_{\alpha \beta} \Big) c_{j\beta}, \label{eq:Hrel}
\end{equation}
where $\bm{\sigma}$ is the Pauli vector. This general Hamiltonian contains the regular hopping terms $t_{ij}$, the molecular spin-orbit coupling $\hat{H}_{SMO}$ (Eq. \ref{eq:Hsoc}) discussed above, as well as inter-molecular spin-orbit coupling terms. Table I gives the single-electron coupling terms produced from the four component relativistic calculation ($t^{rel}$ and $\bm{\lambda}$), and the previously-determind scalar-relativistic equivalents ($t^{sca}$) \cite{jacko15a}.
The first principles Hamiltonian can be expressed as
\begin{equation} 
\hat{H}_{rel} \equiv \hat{H}_0 + \hat{H}_{SMO} + \hat{H}_{SO}^{inter},
\end{equation}
where $\hat{H}_0$ contains the usual tight-binding hopping $t_{ij}$, $\hat{H}_{SMO}$ is the spin-molecular orbital coupling (for the $\mathcal{C}_3$ case), and $\hat{H}_{SO}^{inter}$ contains all of the inter-molecular spin-orbit coupling effects.

\begin{table}[htbp]\label{tab:ts}
	\centering
		\begin{tabular}{|c| r |r |rrr|c|} \hline  
		& $t^{sca}_{ij}$ & $t^{rel}_{ij}$ & $\lambda^{x}_{ij}$ & $\lambda^{y}_{ij}$ & $\lambda^{z}_{ij}$ & ${\bf R}_{ij} = r_j - r_i$    \\ \hline
			$\mu$ 			&-50.195 		&-50.389  & - & - & - & -               															\\ \hline
			$t_k$ 			& 59.692 		& 59.704 	& -0.880 & 0.517 & 1.417 & ${\bf r}_2 -{\bf r}_1$ 					 \\ \hline % 1,2
			$t_g$ 			& 47.112 		& 47.081 	& 0 & 0 & 0 &              ${\bf r}_4 -{\bf r}_1$ 						\\ % 1,4
			$t_g^{-}$ 	&  7.401 		&  7.403 	& 0 & 0 & 0 &              ${\bf r}_4 -{\bf r}_1 + {\bf r}_z$  \\ % 1,4
			$t_z$ 			& 40.851 		& 40.810 	& -0.350 & 0.165 & 0.042 & ${\bf r}_z$ 												  \\ % 1,1
			$t_k^{+}$ 	&  5.326 		&  5.332 	& -0.087 & 0.356 & 0.339 & ${\bf r}_2 -{\bf r}_1 + {\bf r}_z$ 	 \\ % 1,2
			$t_k^{-}$ 	&  5.083 		&  5.091 	& -0.380 &-0.173 & 0.416 & ${\bf r}_2 -{\bf r}_1 - {\bf r}_z$ 	  \\ % 1,2
			$t_h$ 			& -7.565 		& -7.566	& -0.288 & 0.585 &-0.178 & ${\bf r}_5 -{\bf r}_1$ 							   \\ %  1,5
			$t_h^{+}$ 	& 22.882 		& 22.862	&  0.094 & 0.268 & 0.080 & ${\bf r}_5 -{\bf r}_1 + {\bf r}_z$ 	    \\ \hline % 1,5
		\end{tabular}
	\caption{List of $t$ and $\bm{\lambda} = (\lambda^{x},\lambda^{y},\lambda^{z})$ parameters in meV, ordered by $|{\bf R}_{ij}|$, for $|{\bf R}_{ij}| < 20$ \AA, computed from scalar ($sca$) and four component ($rel$) relativistic DFT Wannier overlaps. ${\bf R}_{ij}$ is an example of the path traversed by a hop $t_{ij}$; interactions that are equivalent under $\mathcal{C}_3$ are related by a $\mathcal{C}_3$ rotation of  $\bm{\lambda}$, and those related by inversion are necessarily the same ($\bm{\lambda}$ is a pseudovector). ${\bf r_i}$ labels the origin of the $i$th WO as labeled in Fig. 1 (see Ref \cite{jacko15a} for more details). $\bf{r}_z$ is the interlayer lattice vector. The spin quantisation axis is parallel to $\bf{r}_z$, which is also the $\mathcal{C}_3$ rotation axis of the molecules.}
	\label{tl}
\end{table}

In the scalar relativistic calculation, there are no terms which can cause spin flips or can distinguish between spin up and down; $\bm{\lambda}_{ij} = 0$.
Once we include the effects of relativity, these effects (and therefore $\bm{\lambda}_{ij}$) can be finite.
Note that the $t_{ij}$ change very little with the inclusion of relativistic effects (cf. Table I). 
We define the action of the inversion operator, $\mathcal{I}$, as $\mathcal{I} c^\dagger_{i\alpha} \mathcal{I}^{-1} = c^\dagger_{i+3 \alpha}$ with $i+6 = i$, and sites labeled as in Fig. \ref{fig:molyview}. 
Inversion has a trivial effect on $\hat{H}_{SMO}$, since both $\bm{L}_{mol}$ and $\bm{S}$ are pseudovectors, $\lambda^\alpha_{ij} = \lambda^\alpha_{i+3 \, j+3}$. 
Rotation around the $\mathcal{C}_3$ axis mixes the $x$ and $y$ components of $\bm{\lambda}$, while leaving $\lambda^z$ unchanged; 
$\begin{pmatrix}\lambda^x_{i+1 \, j+1}\\\lambda^y_{i+1 \, j+1}\end{pmatrix} = R_z(2\pi/3) \begin{pmatrix}\lambda^x_{ij}\\\lambda^y_{ij}\end{pmatrix}$
, where $R_z(2\pi/3)$ is the $\mathcal{C}_3$ rotation matrix rotating about the $z$-axis.
These two operations are sufficent to reconstruct the enitre Hamiltonian from the parameters given in Table \ref{tl}.

\begin{figure*}
\begin{center}
\includegraphics[width=1.2\columnwidth]{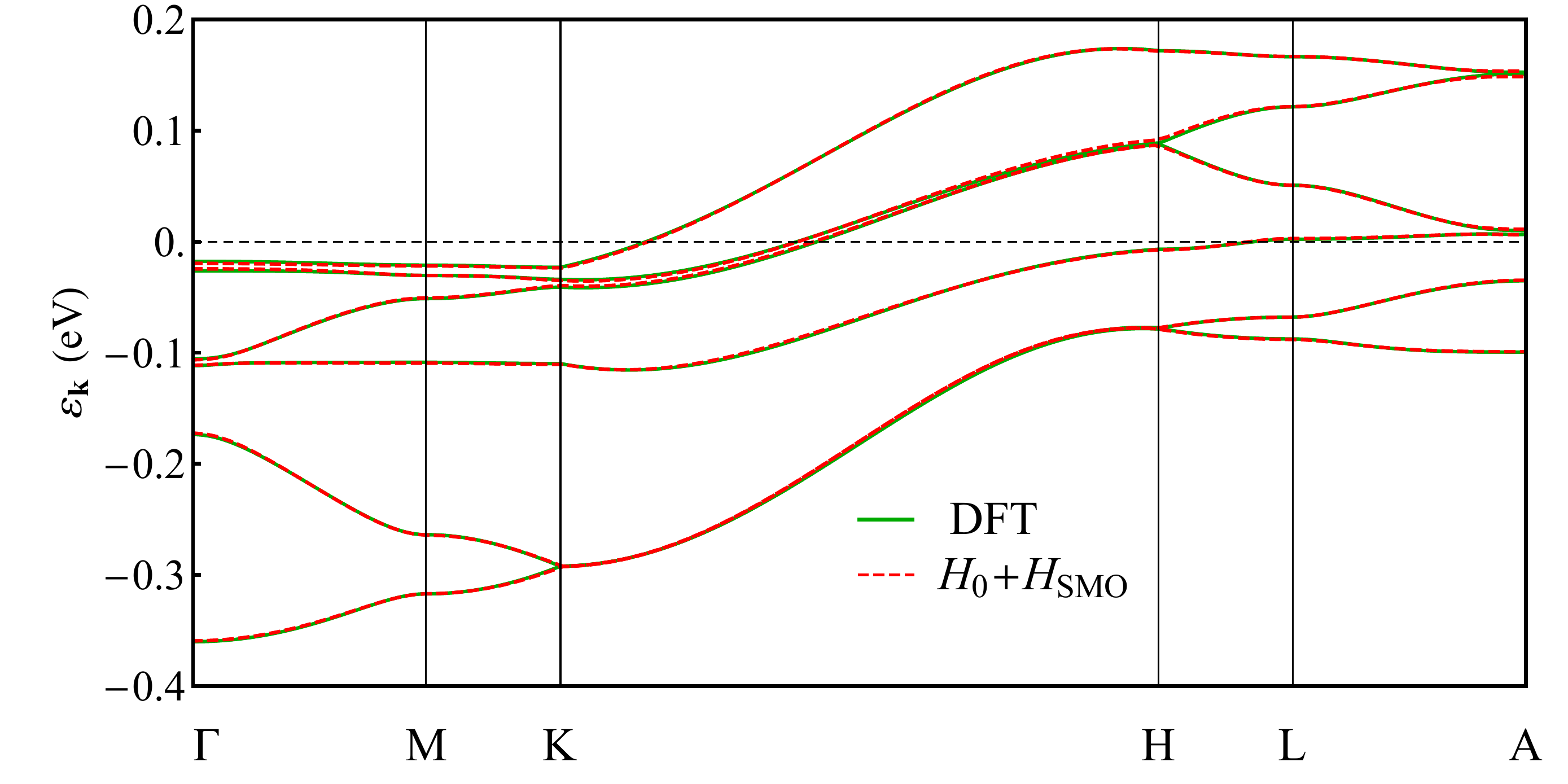} 
 \includegraphics[width=0.63\columnwidth]{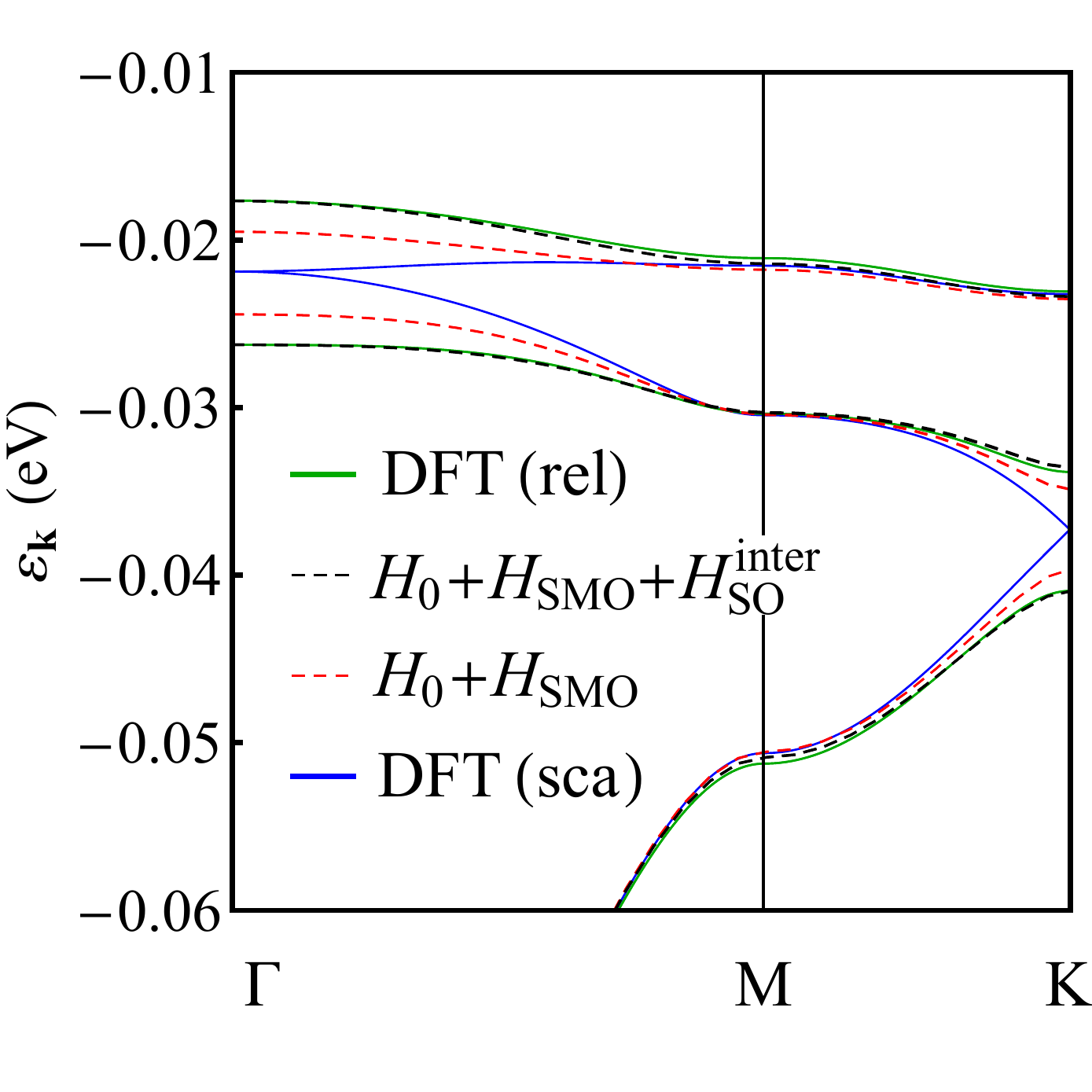} 
\caption{Comparison of four component DFT (green) with our model including $\mathcal{C}_3$ spin-orbit coupling (dashed red), showing very good agreement; on the right the same data is shown in a narrower energy window, with the addition of the scalar relativistic DFT (solid blue), and the model including all computed SO terms (dashed black) up to a cut-off radius of 35 \AA, which shows excellent agreement. Spin-orbit coupling lifts the degeneracy at the Dirac and $\Gamma$ points. This first principles parameterization of $\hat{H}_{SMO}$ captures all of the qualitatives features introduced by relativistic effects in a very simple model. Including inter-molecular SOC up to 35 {\AA} gives excellent agreement with the DFT.
$\hat{H}_{SMO}$ is parameterised from the Wannier overlaps, giving $\lambda_k^z = 4.91$ meV $ = 0.082 t_k$, $\lambda_k^{xy} = 2.50$ meV $ = 0.042 t_k$. $\hat{H}_0$ is the first principles tight-binding model produced from scalar relativistic DFT with $r_c = 35$ \AA, and is in excellent agreement with the DFT electronic structure\cite{jacko15a}.}\label{fig:SOCcompare}
\end{center}
\end{figure*}

\section{Spin-orbit coupling in {\moly}}
We compare the parameters found here to the form of Eq. \ref{eq:Hsoc}.
To do so we transform the spin-orbit coupling operator into the basis of three real-space sites, finding
\begin{eqnarray}\label{eq:socrealspace}
\hat{H}_{SMO} &=& \sum_{j,l=1}^{3} i \Bigg \{ \sin \left[ \phi (j-l)\right] \frac{\lambda_z}{3} \left( \hat{c}_{j\uparrow}^\dagger \hat{c}_{l\uparrow} - \hat{c}_{j\downarrow}^\dagger \hat{c}_{l\downarrow} \right)  \\ \nonumber
& &\phantom{H} + \frac{\sqrt{2} \lambda_{xy}}{3} \left[e^{i \phi j} - e^{i \phi l} \right]  \Big( \hat{c}_{j\uparrow}^\dagger \hat{c}_{l\downarrow} - h.c. \Big) \Bigg \},  \\ \nonumber
\end{eqnarray}
with $\phi = 2 \pi / 3$.
Comparing the Wannier overlaps in Eq. \ref{eq:Hrel} with this form, we find that $\lambda_k^z = 4.91$ meV $ = 0.082 t_k$; $\lambda_k^{xy} = 2.50$ meV $ = 0.042 t_k$ (comparing $\hat{H}_{rel}$ with $\hat{H}_{SMO}$, we see that $\lambda_k^{xy} = \sqrt{6} \sqrt{(\lambda_k^{x})^2 + (\lambda_k^{y})^2} $).
It is interesting to note that this system is quite far from the spherically symmetric case $\lambda_k^z = \lambda_k^{xy}$; reflecting the planarity of the molecule.

The pair of molecules in the unit cell are related by an inversion symmetry, and this has important consequences on the spin-orbit coupling. 
Consider the spin-orbit coupling between a pair of sites related by a $\mathcal{C}_i$ (inversion) symmetry. With the aid of the double group table of $\tilde{\mathcal{C}}_i$, Table II, one can show that there is no allowed spin-orbit coupling contribution along the (inversion symmetric) bond connecting them.
For spinless fermions, there are two possible single particle wavefunctions on the pair of sites related by inversion symmetry, sites 1 and 4 (cf. Fig. 1), connected by the $t_g$ bond; the bonding wavefunction, $\ket{b} = \frac{1}{\sqrt{2}}(\hat{c}^\dagger_{1} + \hat{c}^\dagger_{4})\ket{0}$, which is even and so belongs to the $A_g$ irrep, and the anti-bonding wavefunction $\ket{a} = \frac{1}{\sqrt{2}}(\hat{c}^\dagger_{1} - \hat{c}^\dagger_{4})\ket{0}$ which is odd, and so belongs to $A_u$ (and where $\hat{c}^\dagger_{i}$ creates a fermion on site $i$). 
Considering only the spin component, both $\ket{\uparrow}$ and $\ket{\downarrow}$ belong to $A_{1/2, g}$. 
Only Hamiltonian matrix elements whose symmetry contains the trivial irrep $A_g$ can be non-zero. Any term $\bra{a, \sigma} \hat{H} \ket{b, \sigma'}$  has symmetry $A_{u} \otimes A_{1/2, g} \otimes A_g \otimes A_{g} \otimes A_{1/2, g} = A_u$ since $A_{1/2, g} \otimes A_{1/2, g} = A_g$, $A_{g}\otimes A_{u} = A_u$, and so must be exactly zero. 
Thus, the bonding and anti-bonding sectors are not coupled by SOC. 
Additionally, the $\ket{\sigma}$ and $\ket{\bar{\sigma}}$ states are related by time-reversal symmetry and must form a Kramers doublet. Thus for the two spin states in each sector to remain degenerate they cannot couple to each other.
\begin{table}[t]
	\centering
	\begin{tabular}{l|llll|l}
		$\tilde{\mathcal{C}}_i$ & $E$ & $I$ & $\bar{E}$ & $\bar{I}$ &   \\ \hline 
		$A_g$ & 1 & 1 & 1 & 1 & $\ket{b}$\\ 
		$A_u$ & 1 & -1 & 1 & -1 & $\ket{a}$\\ 
		\hline 
		$A_{1/2, g}$ & -1 & 1 & 1 & -1 & $\ket{\uparrow}$,$\ket{\downarrow}$\\ 
		$A_{1/2, u}$ & 1 & -1 & -1 & 1 & \\ 
		\end{tabular}
	\caption{Character table for  the double group $\tilde{\mathcal{C}}_i$; $E$ is the identity operation, $I$ is the inversion operation, and $\bar{\chi}$ is the group operation $\chi$ plus an additional $C_1$ rotation, i.e. a rotation by $2 \pi$. The $A_x$ are the four irreducible representations (irreps) of $\tilde{\mathcal{C}}_i$, and the characters indicate how states belonging to those irreps transform under the group operations (for example, a wavefunction in $A_u$ changes sign under inversion).
	Representations for bosonic states are given `above the line', while fermionic states are represented below the line. For a more complete explanation of character tables and group theory, see for example \cite{tinkham}. The right-most column shows how example states (bonding ($\ket{b}$), anti-bonding ($\ket{a}$), and spin-$\frac{1}{2}$'s) transform in this group.} 
	\label{tab:ci}
\end{table}
We see this in the DFT results as the spin orbit couplings along the $g$ and $g^{-}$ bonds are precisely zero, $\bm{\lambda}_g = 0$ and $\bm{\lambda}_g^{-} = 0$, as both of these couplings connect sites related by inversion symmetry.

Fig. \ref{fig:SOCcompare} shows the four component DFT band structure as compared to the model with $\hat{H}_{SMO}$ parameterised from the Wannier overlaps. The degeneracies are lifted at the $K$ (Dirac) and $\Gamma$ points. 
The full tight-binding model, Eq. \ref{eq:Hrel}, reproduces the fine details of the four component DFT. We stress that there is no fitting in determining the effective parameters: $t_{ij}$ and ${\bm \lambda}_{ij}$; they are determined directly from the matrix elements between Wannier orbitals. The full parametrization is given in  Table I.
Nevertheless we also find that the simple SMO Hamiltonian, Eq. \ref{eq:Hsoc}, reproduces the essential physics and even provides reasonable quantitative agreement with the four component DFT.

The structure of {\moly} is, as previously discussed, well represented by stacked layers of 2D kagomene sheets.
The effects of spin-orbit coupling in 2D systems is a field of ongoing interest \cite{kane05,kane05b,min06,ruegg10,xiao12}.
In 2D systems, the gradients of the potential in the plane are expected to be quite different from those perpendicular to the plane, so these are considered separately $\bm{\nabla}V = \bm{\nabla}V_{xy} + \bm{\nabla}V_z$. 
As such, it is natural to split $\bm{\nabla}V \times \bm{p}$ into a `Rashba' term, $\bm{\nabla}V_z \times \bm{p}$, that depends only on the gradient of the potential perpendicular to the 2D plane, and an `SO' term,  $\bm{\nabla}V_{xy} \times \bm{p}$, that contains the in-plane gradients of the potential. In 2D, $\bm{p} = (p_x, p_y, 0)$, so the Rashba term can cause spin-flips (c.f. $\lambda_{xy}$), and the SO term is a spin-dependent hopping that does not flip spins (c.f. $\lambda_z$).
A previously studied relativistic model of {\moly}'s lattice at $\frac{2}{3}$ filling is found to have a quantum spin-Hall (QSH) insulating phase for the parameter values we find here for {\moly} \cite{ruegg10}. It is worth noting that, in contrast to the previous work\cite{ruegg10}, our orbital angular momentum model predicts contributions of both $\lambda^{xy}$ and $\lambda^z$ on the clusters, and both of these terms are found to be non-zero in the relativistic DFT. It is unclear how these and other additional relativistic contributions will modify the found QSH ground state. Nevertheless mono-layer {\moly} may well demonstrate such a quantum spin-Hall insulating ground state.

It has been seen that triangular clusters coupled as {\moly} is in the $x-y$ plane and stacked in the $z$ direction lead to a quasi-1D spin chain known to have a topological Haldane phase ground state, consistent with experimental evidence \cite{noursehaldane, janani14a}. It has further been argued that SMOC can drive a phase transition from a topological (Haldane) phase to a trivial phase in such chains \cite{merino16moly}. 
However, the effects of inter-cluster spin-orbit coupling on this topological ground state has not yet been considered. 
As Table I shows, the intra-cluster spin-orbit coupling ($\bm{\lambda}_k$) is of the same order as the intra-cluster terms (all other $\bm{\lambda}_{ij}$). With this detailed model, one can now investigate the effects of these terms on the stability of the Haldane phase.

\section{Discussion} 

Here we have demonstrated a powerful application of the Wannier orbital construction technique, the computation of first principles spin-orbit coupling parameters in complex molecular materials. Unlike atoms and atomic solids, the form of spin-orbit coupling in molecular crystals is not the usual $\bm{L} \cdot \bm{S}$. In these systems, relativistic effects are known to be important, but there has not been a robust strategy for incorporating them consistently. By using the Wannier parameterization in a relativistic four component DFT computation, one can determine the relativistic contributions and incorporate them into further modelling.

In {\moly}, the leading relativistic effects are well described by a coupling between the spin$-\frac{1}{2}$ electron and emergent spin$-1$ molecular orbital angular momentum states. Our first principles parameterization shows us that there are additional SOC terms coupling molecules together. 
These intermolecular SOC contributions have significant effects on, for example, the magnitude of the gap between bands at the $\Gamma$ and $K$ high-symmetry points. 
We also found that along inversion-symmetric bonds there are no relativistic contributions to the single-electron model, and gave a group-theoretic explanation for this observation. While the magnitude of the spin-orbit coupling observed here is small ($|\bm{\lambda}_k|/t_k \sim 10$\% ), the chemical flexibility of molecular crystals allows us to tune the magnitude of the spin-orbit coupling, intra-molecular hopping, and inter-molecular hopping \cite{kanoda11,powell06}. For example, in a tungsten analogue of {\moly}, the spin-orbit coupling could scale up by as much as a factor of $(Z_W/Z_{Mo})^4 = (74/42)^4 \sim 10$. At the same time, one could consider modifying the dmit ligands to reduce the inter-molecular hopping. The anisotropic exchange interactions caused by SOC means that with this kind of control one could realise compass models such as the Kitaev model in this class of molecular crystals.

\section*{Acknowledgments}
This work was supported by the Australian Research Council through grants FT130100161, DP130100757, DP160100060 and LE120100181. J.M. acknowledges Financial support from (MAT2015-66128-R) MINECO/FEDER, UE. Density functional calculations were performed with resources from the National Computational Infrastructure (NCI), which is supported by the Australian Government.

\end{document}